\documentclass{eas}
\usepackage{graphicx}
\begin{document}

\title{Red Supergiants in the Local Group} 
\author{Emily M. Levesque}\address{CASA, Department of Astrophysical and Planetary Sciences, University of Colorado 389-UCB, Boulder, CO 80309, USA; Einstein Fellow} 
\begin{abstract}
Galaxies in the Local Group span a factor of 15 in metallicity, ranging from the super-solar M31 to the Wolf-Lundmark-Melotte (WLM) galaxy, which is the lowest-metallicity (0.1Z$_{\odot}$) Local Group galaxy currently forming stars. Studies of massive star populations across this broad range of environments have revealed important metallicity-dependent evolutionary trends, allowing us to test the accuracy of stellar evolutionary tracks at these metallicities for the first time. The RSG population is particularly valuable as a key mass-losing phase of moderately massive stars and a source of core-collapse supernova progenitors. By reviewing recent work on the RSG populations in the Local Group, we are able to quantify limits on these stars' effective temperatures and masses and probe the relationship between RSG mass loss behaviors and host environments. Extragalactic surveys of RSGs have also revealed several unusual RSGs that display signs of unusual spectral variability and dust production, traits that may potentially also correlate with the stars' host environments. I will present some of the latest work that has progressed our understanding of RSGs in the Local Group, and consider the many new questions posed by our ever-evolving picture of these stars.
\end{abstract}
\maketitle
\section{Introduction} 
Red supergiants (RSGs) are a critical helium-burning evolutionary stage in the lives of moderately massive (10-25$M_{\odot}$) stars. Their large physical size, extended atmospheres, and cool effective temperatures ($T_{\rm eff}$s) mark RSGs as a unique and extreme phase of massive stellar evolution. In addition, the high bolometric luminosities ($M_{\rm bol}$s) of RSGs make them excellent observational sources within our own galaxy as well as ideal targets for extragalactic stellar astrophysics, extending the study of individuals stars out to, and potentially even beyond, the Local Group.

Recent observations of Galactic RSGs helped to resolve a long-standing disagreement between these stars' physical properties and the predictions of evolutionary models. Massey (2003) and Massey \& Olsen (2003) both noted that the $T_{\rm eff}$s and bolometric luminosities ($M_{\rm bol}$) determined for RSGs in Humphreys (1978) and Humphreys \& McElroy (1984) were too cool and too luminous to agree with the predictions of the Geneva evolutionary tracks (Schaller et al.\ 1992, Meynet et al.\ 1994). This was originally assumed to be a deficiency in the tracks, given that RSG atmospheres are extremely hard to model. However, Levesque et al.\ (2005) obtained optical spectrophotometry for a sample of 74 RSGs in the Milky Way. We determined these stars' physics properties by combining our observations with the new generation of MARCS stellar atmospheres, which included greatly-improved treatments of oxygen-rich molecule opacities (Plez 2003, Gustaffson et al.\ 2003, Gustaffson et al.\ 2008). The $T_{\rm eff}$s determined with these observations were markedly warmer than previous results, bringing Galactic RSGs into excellent agreement with evolutionary models.

Since these initial Galactic observations, the study of RSGs has been extended to extragalactic samples, using the two-color selection method of Massey (1998) to eliminate foreground dwarfs and robustly identify RSG members of nearby galaxies. With observations of RSG populations spanning a wide range of metallicities, we are able to examine the abundance-dependent changes on these stars' physical properties. The effective temperatures, luminosities, mass loss properties, and variability of the diverse Local Group RSG populations all hold important information about metallicity-dependent massive star evolution.

\section{RSGs and Metallicity} 
Magellanic Cloud RSGs originally showed the same discrepancy with the evolutionary models that was observed in the Milky Way - evolutionary tracks did not extend to high enough luminosities or cool enough temperatures to agree with the RSG samples of Elias et al.\ (1985) and Massey \& Olsen (2003). Following our redetermination of $T_{\rm eff}$s for the Galactic RSG sample, we performed a similar analysis on optical spectrophotometry of 36 RSGs in the Large Magellanic Cloud (LMC) and 37 Small Magellanic Cloud (SMC) RSGs (Levesque et al.\ 2006). The new $T_{\rm eff}$ and $M_{\rm bol}$ values determined for these stars produced similar results to our work in the Milky Way, largely resolving the disagreement between RSGs and the evolutionary tracks in the LMC. However, there was an improved but not wholly satisfactory agreement with the evolutionary tracks for the SMC stars, with this sample showing a considerably larger spread in $T_{\rm eff}$ for a given value of $M_{\rm bol}$ as compared to the higher-metallicity samples. It is important to note that this larger spread in $T_{\rm eff}$ and the persistent disagreement with evolutionary models is not entirely surprising due to the expected enhancement of rotational mixing effects in stars at these lower metallicities (e.g. Maeder \& Meynet 2001), and it is possible that new stellar evolutionary models that include treatments of rotation effects at these lower metallicities will show better agreement with the RSG sample (see, for example, Ekstr\"{o}m et al.\ 2012).

Most recently, the Local Group dwarfs NGC 6822 and Wolf-Lundmark-Melotte (WLM) were also both found to have substantial RSG populations in Massey et al.\ (2007a). NGC 6822 has a metallicity of $Z = 0.4Z_{\odot}$), intermediate between the LMC and SMC, while with $Z = 0.1Z_{\odot}$ WLM is the lowest-metallicity Local Group galaxy currently forming stars. Levesque \& Massey (2012) acquired spectroscopy of 16 RSGs in NGC 6822 and spectrophotometry of 11 RSGs in WLM, determining spectral types for the RSGs in these samples and comparing them to the Milky Way and Magellanic Cloud results. 

The effect of metallicity on RSGs in the Local Group is illustrated in Figure 1. Originally noted by Elias et al.\ (1985) and further explored by Massey \& Olsen (2003), the average spectral type of an RSG population shifts to earlier subtypes with decreasing metallicity. The average RSG subtype in the Milky Way ($Z = Z_{\odot}$) is M2 I, but this average shifts to M1 I in the LMC ($Z =  0.5Z_{\odot}$), K5-7 I in the SMC ($Z = 0.02Z_{\odot}$), and K0-1 in WLM ($Z = 0.1Z_{\odot}$). At a glance, the physical explanation for this progression seems straightforward - RSG spectral types are dependent on the strength of their optical TiO bands, and stars with lower atmospheric abundances and weaker TiO lines will therefore have weaker TiO features and be assigned earlier spectral types. However, this observed shift is also largely attributed to RSG populations in lower-metallicity galaxies having warmer median effective temperatures. This trend is a consequence of a decrease in the number of heavy elements contributing electrons, which corresponds to a decrease in opacity and a subsequent increase in the stars' surface temperature. On the Hertzsprung-Russell diagram this is illustrated as a shift of the Hayashi limit (Hayashi \& Hoshi 1961) to warmer temperatures. Since this limit represents a restriction on how cool (and as a result, how late-type) a RSG can get while still maintaining hydrostatic equilibrium, RSG populations as a whole are warmer in low-metallicity environments.

\begin{figure}
\includegraphics[width=12cm]{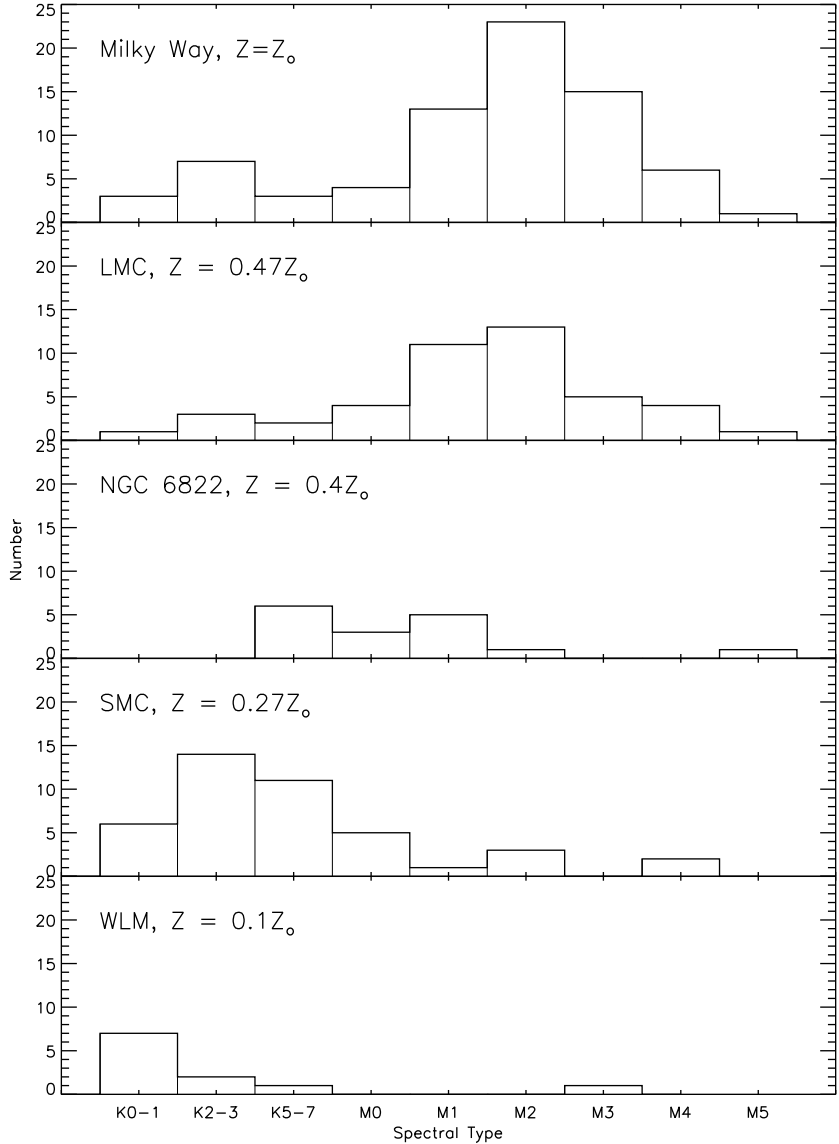}{\centering}
\caption{Adapted from Levesque \& Massey (20120; histograms of RSG spectral subtypes in the Local Group, plotted from top to bottom in order of decreasing host metallicity. Data are from Levesque et al.\ (2005; Milky Way), Levesque et al.\ (2006, 2007) and Massey et al.\ (2007b; Magellanic Clouds), and Levesque \& Massey (2012; NGC 6822 and WLM). There is a clear progression of the dominant spectral subtype towards earlier types (corresponding to weaker TiO features) at lower metallicities, in agreement with previous discussion in Elias et al.\ (1985) and Massey \& Olsen (2003).}
\end{figure}

Metallicity's effects are expected to extend beyond changes in these stars' $T_{\rm eff}$s. There is an expected relation between the maximum luminosities ($L_{\rm max}$) of RSGs and their metallicities, with  lower-metallicity RSGs having higher luminosities (Massey 1998). This dependence should arise from abundance-dependent mass loss effects; while a high-metallicity massive star may immediately become a Wolf-Rayet star when it begins helium burning, a low-metallicity star of the same mass should evolve through an intermediate, and potentially even terminal, RSG phase as a result of shedding its outer layers at a much lower rate.

These predictions, however, are not supported by extragalactic RSG observations; Massey et al.\ (2009) compare the maximum $M_{\rm bol}$ determined for Milky Way, Magellanic Cloud, and M31 RSGs and find that all three environments share a common maximum luminosity of log($L/L_{\odot}$) $\sim$ 5.2-5.3, showing no change in $M_{\rm bol}$ with abundance even across the $\sim$0.9 dex metallicity spread between the SMC and M31. However, this could be a limitation of selection techniques for extragalactic RSGs. The metallicity-dependent shift of the Hayashi track complicates color-dependent selection criteria; with warmer temperatures associated with lower metallicities, the late-type massive star population of a metal-poor galaxy is expected to be collectively bluer. As a result, selecting only the reddest massive stars (see, for example, Massey et al.\ 1998, Levesque et al.\ 2006, Levesque \& Massey 2012) excludes the ``orange" or even yellow supergiants that are necessary for effectively sampling the high-luminosity end of the evolved massive star population in these galaxies.

\section{Unusual Red Supergiants} 
\subsection{Dust-Enshrouded RSGs}
Massive stars lose more than half of their mass after they evolve off of the main sequence (e.g., Stothers \& Chin 1996), and for 10-25M$_{\odot}$ stars much of this mass loss occurs during the RSG phase. Most RSGs undergo some amount of sporadic mass loss that is evidenced by the presence of circumstellar dust shells (e.g. Stencel et al.\ 1988, 1989; Danchi et al.\ 1994; Massey et al.\ 2005). However, a subset of RSGs show evidence of extreme mass loss and dust production. These stars are characterized by thick asymmetric circumstellar dust nebulae (making them extremely bright IR sources) and the presence of strong OH, SiO, and H$_2$O masers produced in the outflowing material generated by high mass loss episodes. While some maser activity in late-type RSGs is not necessarily unusual (e.g. Figer et al.\ 2006, Davies et al.\ 2008, Verheyen et al.\ 2012), the strong OH masers and substantial dust nebulae associated with these extreme RSGs - often referred to as ``OH/IR stars" - definitively sets them apart as distinct from the general population.

A handful of Galactic OH/IR stars have been identified and are well-studied, including VY CMa, VX Sgr, S Per, and NML Cyg (e.g. Smith et al.\ 2001; Humphreys et al.\ 2005, 2007; Schuster et al.\ 2006; Jones et al.\ 2007). Photometric observations of these stars are quite extensive, with data spanning $\sim$50-100 years and showing excellent evidence of their substantial circumstellar envelopes. Spectrophotometry of these stars is more challenging, and must accommodate the effects of line-of-sight dust; as a result, determining $T_{\rm eff}$ and $M_{\rm bol}$ for these stars is challenging (see Massey et al.\ 2006a, Choi et al.\ 2008).

WOH G64, an unusual RSG in the Magellanic Clouds, is currently the most robust example of an extragalactic OH/IR star. This RSG is a strong SiO, H$_2$O, and OH maser source (e.g. Wood et al.\ 1986, van Loon et al.\ 2001) and shows evidence of extreme mass loss and a thick toroidal dust envelope (e.g. Elias et al.\ 1986, Marshall et al.\ 2004, Ohnaka et al.\ 2008). Optical spectra of this RSG correspond to a spectral type of M5 I, unusually late for the LMC and corresponding to an extremely cold temperature of 3400 K that places it beyond the Hayashi limit on the Hertzsprung-Russell diagram. Finally, spectra of WOH G64 also show evidence of a nebular emission line spectrum, with an unexplained excitation mechanism that could be attributable to shock heating or ionization by a hot binary companion (Levesque et al.\ 2009). With the current small sample size of OH/IR stars, it is unclear whether these features are associated with WOH G64's mass loss behavior or indicative of additional abnormalities in this RSG.

\subsection{Variable RSGs} 
As described in Section 2, the Hayashi limit is observed to shift to warmer temperatures at lower metallicities. This limit restricts how cold, and hence how late-type, RSGs can be in a particular environment. The expected result of this metallicity-dependent limit is a lack of cold RSGs in lower-metallicity galaxies such as the Magellanic Clouds, NGC 6822, and WLM. 

Despite this expectation, Figure 1 clearly demonstrates that a small number of aberrantly late-type RSGs are still present in all of these low-metallicity environments. These outliers were first examined in Levesque et al.\ (2007) and Massey et al.\ (2007b). With multiple observations of these stars, we found large discrepancies in the spectral subtypes assigned to them over $\sim$1 year timescales. The most dramatic example of this variation is HV 11423, a SMC RSG, whose spectral type has varied from K0-1 I in 2004 to M4 I in 2005 and an intermediate type of M2 I in 2010, corresponding to $T_{\rm eff}$ measurements of 4300 K, 3500 K, and 3625 K respectively (Massey et al.\ 2007b, Levesque \& Massey 2012). HV 11423 also displays abnormally high variability in $V$ (well in excess of the $\sim$1 mag variations that are typical of most RSGs; Josselin et al.\ 2000, Levesque et al.\ 2007), $M_{\rm bol}$ and $V$-band reddening. While HV 11423 is the most extreme case, all four of the Magellanic Cloud variables share the same behavioral template: when they appear warmer they are also brighter, dustier, and more luminous, and undergo these variations on a timescale of months.

The RSGs J194453.46-144552.6 (in NGC 6822) and J000158.14-152332.2 (in WLM) also have aberrantly late spectra types, although without a second epoch of observations it is unclear whether these RSGs are members of the this variable class or late-type dust-enshrouded RSGs like WOH G64. Another RSG in NGC 6822, J194449.96-144333.5, does show conclusive evidence of variable behavior; the star was found to have a spectral type of M2.5 I in observations from 1997 (Massey 1998), while in Levesque \& Massey (2012) the star was captured in in earlier-type state as a K5 I.

It is currently believed that these variable stars are undergoing an unstable, and likely short-lived, evolutionary phase. However it is unclear whether the changes in these observed properties are indicative of physical changes in the star's atmosphere or merely a consequence of sporadic mass-loss episodes and resulting changes in the stars' optical depths and apparent spectra (see also Levesque 2010).

\section{The Future of Extragalactic RSG Studies} 
In recent years we have begun to explore the RSG population of the Local Group. Many nearby galaxies contain rich populations of RSGs that still remain to be studied. Massey et al.\ (2009) presented optical spectrophotometry of 16 RSGs in M31, focusing on a small sample of stars that represented the galaxy's ``average" spectral subtype of M2-3 I; however, the full sample of RSGs in M31 is much larger (see Massey et al.\ 2006b). A dedicated spectroscopic survey of these RSGs would dramatically increase our sample size of high-metallicity RSGs, and could even be used in conjunction with stellar evolution models to help answer outstanding questions about the metallicity of M31 (see further discussion in Massey et al.\ 2009). A number of additional RSGs have also been identified in the LMC (Neugent et al.\ 2012), as well as in M33 (Drout et al.\ 2012) and Sextans B and A (Massey et al.\ 2007a). All of these stars are prime candidates for follow-up spectroscopic observations, ultimately leading to an extensive sample of extragalactic RSGs that span a broad range of environments and can serve as important tests of stellar evolutionary theory.

Recent studies of Galactic and Magellanic Cloud RSGs have highlighted the importance of both optical {\it and} infrared observations of these cool stars. Levesque et al.\ (2005, 2006) and Massey et al.\ (2009) consider the value of ($V-K$) colors and $K$-band luminosities in determining the physical properties of RSGs. van Belle et al.\ (2009) compares properties determined from infrared interferometry to those determined based on spectral types and ($V-K$) colors. Finally, Davies et al.\ (2013) consider the differences in $T_{\rm eff}$s determined from optical and infrared spectra and the resulting implications for the importance of 3D models when studying the extended atmospheres of these stars.

In addition to using RSG physical properties as direct tests of stellar evolutionary models, RSG populations have a number of other important applications. The blue-to-red supergiant ratio and RSG-to-Wolf-Rayet ratio of galaxies both hold important information about stellar evolution models, in particular their treatment of mass loss and other abundance-related evolutionary effects. RSGs have also been identified as Type II supernova progenitors as well as a potential intermediate stage of stellar evolution for the more massive progenitors of Type Ibc supernovae (for a more detailed discussion see Smartt 2009).

It is important to note that RSGs are only one of several key post-main-sequence evolutionary phases. Further exploring the relative population ratios of RSGs, or the fraction of supernova progenitors that they encompass, will also benefit from similar studies of other evolved massive stars such as Wolf-Rayet stars and yellow supergiants. Ideally, by building a library of well-studied evolved massive stars that includes RSGs, we will be able to refine stellar evolutionary models, improve our understanding of core-collapse supernova, and place these stars in a population-wide context that spans a broad range of environmental conditions. With these data in hand, we can also make important strides in our understanding of stellar evolution that will prepare us for extending this work to galaxies star beyond the Local Group in the next decade as facilities such as the James Webb Space Telescope and the Extremely Large Telescopes come online. \\

The author gratefully acknowledges her collaborators in this work on RSGs, in particular Philip Massey, Knut Olsen, Bertrand Plez, George Meynet, Andre Maeder, Nidia Morrell, Eric Josselin, Maria Drout, and Kathryn Neugent. Support for this work was provided in part by NASA through Einstein Postdoctoral Fellowship grant number PF0-110075 awarded by the Chandra X-ray Center, which is operated by the Smithsonian Astrophysical Observatory for NASA under contract NAS8-03060.


\end{document}